\documentclass[aps]{revtex4}
\def\lta{~\raise.4ex\hbox{$<$}\llap{\lower.6ex\hbox{$\sim$}}~}
\def\gta{~\raise.4ex\hbox{$>$}\llap{\lower.6ex\hbox{$\sim$}}~}

\begin{document} \input psfig.sty 

\title{Self-organization in a simple model of adaptive agents playing 
$2\times2$ games with arbitrary payoff matrices}
\author{H. Fort$^{1}$ and S. Viola$^{2}$}

\address{$^{1}$Instituto de F\'{\i}sica, Facultad de Ciencias, Universidad de la
Rep\'ublica, Igu\'a 4225, 11400 Montevideo, Uruguay\\
$^{2}$Instituto de F\'{\i}sica, Facultad de Ingenier\'{\i}a, Universidad de la
Rep\'ublica, Julio Herrera y Reissig 565, 11300 Montevideo, Uruguay}

\begin{abstract}

We analyze, both analytically and numerically, the self-organization of a system of "selfish" adaptive agents playing an arbitrary iterated pairwise game ( defined by a 2$\times$2 payoff matrix).
Examples of possible games to play are: the {\it Prisoner's Dilemma} (PD) game, 
the {\it chicken} game, the {\it hero} game, etc.  
The agents have no memory, use strategies not based on direct 
reciprocity nor 'tags' and are chosen at random {\it i.e.} geographical vicinity is neglected. They can play two possible strategies: cooperate (C) or defect (D). The players measure their success by comparing their utilities 
with an estimate for the expected benefits and update their strategy following a simple rule. 

Two versions of the model are studied: 1) the deterministic version 
(the agents are either in definite states C or D) and
2) the stochastic version (the agents have a probability $c$ of playing C). 

Using a general Master Equation we compute the equilibrium states into which 
the system self-organizes, characterized by their average 
probability of cooperation $c_{eq}$. Depending on the payoff matrix, 
we show that $c_{eq}$ can take five different values.

We also consider the mixing of agents using two different payoff matrices an show that any value of $c_{eq}$ can be reached by tunning the proportions of agents using each payoff matrix. In particular, this can be used as a way to simulate the effect a fraction $d$ of "antisocial" individuals -incapable of realizing any value to cooperation- on the cooperative regime hold by a population of neutral or "normal" agents.

\end{abstract}

\maketitle

\vspace{2mm}


\vspace{2mm}

\section{Introduction}

Complex systems pervade our daily life. They are difficult to study because 
they don't exhibit simple cause-and-effect relationships and their 
interconnections are not easy to disentangle. 

Game Theory has demonstrated to be a very flexible tool to study complex 
systems. It coalesced in its {\it normal form}\cite{hs88} during the second World War with the work of Von Neumann and Morgenstern \cite{vNM44} who first applied it in Economics. 

Later, in the seventies, it was the turn of Biology mainly with 
the work of J. Maynard-Smith \cite{ms}, who shown that the Game Theory can 
be applied to various problems of evolution, and proposed the concept of 
Evolutionary Stable Strategy (ESS), as an important concept for understanding 
biological phenomena. 
Following rules dictated by game theory to attain an ESS requires neither consciousness nor a brain.
Moreover, a recent experiment found that two variants of a RNA virus seem
to engage in two-player games \cite{tch99}. 

This opens a new perspective, perhaps the dynamic of very simple agents, of the kind we know in Physics, can be modeled  by Game Theory providing an alternative approach to physical problems. 
For instance, energies could be represented as payoffs 
and phenomena like phase transitions understood as many-agents games.   
As a particular application of this line of thought we have seen recently a 
proliferation of papers addressing the issue of {\it quantum games} \cite{meyer99}- 
\cite{lj02b} which might shed light on the hot issue of quantum computing.
Conversely, Physics can be useful to understand the behavior of adaptive agents playing games used to model several complex systems in nature. 
For instance, in some interesting works Szab\'o {\it et al} \cite{st99},\cite{sasd00} applied the sophisticated techniques developed in non-equilibrium statistical physics to spatial evolutionary games. 

The most popular exponent of Game Theory is the {\it Prisoner's Dilemma} (PD) 
game introduced in the early fifties by M. Flood and M. Dresher \cite{f52} 
to model the social behavior of "selfish" individuals - individuals which pursue 
exclusively their own self-benefit. 

The PD game is an example of a $2\times2$ game in normal form: i) there are 2 
players, each confronting 2 choices - to cooperate (C) or to defect (D)-, ii) 
with a $2\times2$ matrix specifying the payoffs of each player for the 4 
possible outcomes: [C,C],[C,D],[D,C] and [D,D]\footnote{[X,Y] means that the 
first player plays X and the second player plays Y (X an Y = C or D ).}
and iii) each player makes his choice without knowing what the other will do. 
A player who plays C gets the "reward" $R$ or the "sucker's payoff" $S$ depending  
if the other player plays C or D respectively, while if he plays D he gets 
the "temptation to defect" $T$ or the "punishment" $P$ depending if the other player 
plays C or D respectively. These four payoffs obey the relations:
\begin{equation}
T>R>P>S,
\label{eq:inequal}
\end{equation}
and
\begin{equation}
2R>S+T.
\label{eq:dilemma2}
\end{equation}
Thus independently of what the other player does, by (\ref{eq:inequal}), 
defection D yields a higher payoff than cooperation C ($T>R$ and $P>S$) and is 
the {\em dominant strategy}. The outcome [D,D] is thus called a Nash 
equilibrium \cite{n51}. 
The dilemma is that if both defect, both do worse 
than if both had cooperated ($P<R$). Condition (\ref{eq:dilemma2}) is required
in order that the average utilities for each agent of a cooperative pair ($R$) 
are greater than the average utilities for a pair exploitative-exploiter 
(($T+S$)/2). 

Changing the rank order of the payoffs - the inequalities (\ref{eq:inequal})- gives rise to different games.
A general taxonomy of $2\times2$ games (one-shot games involving two players 
with two actions each) was constructed by Rapoport and Guyer \cite{rg66}. 
A general $2\times 2$ game is defined by a payoff matrix M$^{RSTP}$ with 
payoffs not necessarily obeying the conditions (\ref{eq:inequal}) or 
(\ref{eq:dilemma2})\footnote{We will maintain the letters $R,S,T$ or $P$ 
to denote the payoffs in order to keep the PD standard notation.}
\begin{equation}
{\mbox M}^{RSTP}=\left(\matrix{(R,R)&(S,T)\cr (T,S)&(P,P) \cr}\right).
\label{eq:payoffmatrix}
\end{equation}
The payoff matrix gives the payoffs for {\em row} actions when confronting
with {\em column} actions.  

Apart from the PD game there are other some well studied games.
For instance, when the damage from mutual defection in the PD is increased 
so that it finally exceeds the damage suffered by being exploited:
\begin{equation} 
T>R>S>P,
\label{eq:chicken}
\end{equation}
the new game is called the {\it chicken} game.
Chicken is named after the car racing game. Two cars drive towards each 
other for an apparent head-on collision. Each 
player can swerve to avoid the crash (cooperate) or keep going (defect). 
This game applies thus to situations such that mutual defection is the worst 
possible outcome (hence an unstable equilibrium).

When the reward of mutual cooperation in the chicken game is decreased so 
that it finally drops below the losses from being exploited:
\begin{equation}
T>S>R>P,
\label{eq:leader}
\end{equation}
it transforms into the {\it leader} game. The name of the game stems 
from the following every day life situation: Two car drivers want to enter a 
crowded one-way road from opposite sides, if a small gap occurs in the line of 
the passing cars, it is preferable that one of them take the lead and enter 
into the gap instead of that both wait until a large gap occurs and allows 
both to enter simultaneously.
  
In fact, every payoff matrix, which at a first glance could seem 
unreasonable from the point of view of selfish individuals, can be applicable 
to describe real life situations in different realms or contexts.    
Furthermore, "unreasonable" payoff matrices can be used by minorities of 
individuals which depart from the "normal" ones (assumed to be neutral) 
for instance, absolutely D individuals incapable of realizing any value to 
cooperation or absolutely C "altruistic" individuals (more on this later).

In one-shot or non repeated games, where each player has a dominant strategy, as in the PD, then generally these strategies will be chosen.
The situation becomes more interesting when the games are played repeatedly. In
these {\it iterated games} players can modify their behavior with time in order 
to maximize their utilities as they play {\it i.e.} they can adopt different 
strategies. 
In order to escape from the non-cooperative Nash equilibrium state of social dilemmas it is 
generally assumed either memory of previous interactions \cite{axel84} or 
features ("tags") permitting cooperators and defectors to distinguish one 
another \cite{ep98}; or spatial structure is required \cite{nm93}.

Recently, it was proposed \cite{PRE03} a simple model of selfish agents without memory of past encounters, without tags and with no spatial structure playing an arbitrary $2\times2$ game, defined by a general payoff matrix like (\ref{eq:payoffmatrix}). 
At a given time $t$, each of the $N_{ag}$ agents, numbered by an index $i$, has a probability $c_i(t)$ of playing C ($1-c_i(t)$ of playing D). 
Then a pair of agents are selected at random to play. All the players use the same measure of success to evaluate if they did well or badly in the game which is based on a comparison of their utilities $U$ with an estimate of the expected 
income $\epsilon$ and the arithmetic mean of payoffs $\mu \equiv (R+S+T+P)/4$.
Next, they update their $c_i(t)$ in consonance, i.e. a player keeps
his $c_i(t)$ if he did well or modifies it if he did badly.

Our long term goal is to study the quantum and statistical versions of this 
model. That is, on one hand to compare the efficiency and properties of 
quantum strategies vs. the classical ones for this model in a spirit similar 
to that of ref. \cite{meyer99}. 
On the other hand, we are also interested in the effect of noise, for instance 
by introducing a Metropolis Monte-Carlo temperature, and the existence of 
power laws in the space of payoffs that parameterize the game, of the type 
found in ref. \cite{st99} and \cite{sasd00}, for a spatial structured version 
of this model. 
Before embarking on the quantum or statistical mechanics of this model, 
the objective in this paper is to complete the study of the simplest
non-spatial M-F version. In particular, to present an analytic derivation 
of the equilibrium states for any payoff matrix {\it i.e.} for an arbitrary 
$2\times2$ game using elemental calculus, both for the deterministic and
stochastic versions. In the first case the calculation is elementary and serves as a guide to the more subtle computation of the stochastic model.
These equilibrium states into which the systems self-organizes, which depend 
on the payoff matrix, are of three types: "universal cooperation" or "all C", 
of intermediate level of cooperation and "universal 
defection" or "all D" with, respectively, $c_{eq}$ = 1.0, 
$0 < c_{eq} < 1.0$ and 0.0. 
We also consider the effect of mixing players using two different payoff 
matrices. Specifically, a payoff matrix producing $c_{eq}$=0.0 and the canonical payoff matrix are used to simulate, respectively, absolutely D or "antisocial" agents and "normal" agents.

\section{The Model}

We consider two versions of the model introduced in ref. \cite{PRE03}.
First, a deterministic version, in which the 
agents are always in definite states either C or D 
{\it i.e.} "black and white" agents without "gray tones".
Nevertheless, 
it is often remarked that this is clearly an over-simplification 
of the behavior of individuals.Indeed, their levels of cooperation 
exhibit a continuous gamma of values. Furthermore, 
completely deterministic algorithms fail to incorporate
the stochastic component of human behavior.
Thus, we consider also a stochastic version, in which the agents only 
have probabilities for playing C. 
In other words, the variable $c_i$, denoting the state or "behavior" of 
the agents, for the deterministic case takes only two values: $c_i$ = 1 
(C) or 0 (D) while for the stochastic case  $c_i$ is a real variable 
$\in [0,1]$. 

The pairs of players are chosen randomly instead of being restricted to 
some neighborhood. The implicit assumptions behind this are that the 
population is sufficiently large and the system connectivity is high. 
In other words, the agents display high mobility or they can experiment 
interactions at a distance (for example electronic transactions, etc.). 
This implies that $N_{ag}$ the number of agents needs to be reasonably large.
For instance, in the simulations presented in this work the population of agents will be fixed to $N_{ag}=1000$.

The update rule for the $c_k$ of the agents is based on comparison 
of their utilities with an estimate. 
The simplest estimate $\epsilon_k$ that agent number $k$ for his  
expected utilities in the game is provided
by the utilities he would made by playing with 
himself \footnote{ One might consider more sophisticated agents which have "good" information (statistics, surveys, etc) from which they can extract 
the average probability of cooperation at "real time" $c(t)$ to
get a better estimate of their expected utilities. 
However, the main results do not differ from the ones obtained with this 
simpler agents}, that is: 
\begin{equation}
\epsilon^{R S T P}_k(t) =  (R-S-T+P) c_k(t)^2 +(S+T-2P) c_k(t) +P,
\label{eq:epsilon}
\end{equation}
where $c_k$ is the probability that in the game the agent k plays C.
From equation (\ref{eq:epsilon}) we see that the estimate for C-agents 
($c_k=1$) $\epsilon_C$ and D-agents ($c_k=0$) $\epsilon_D$ are given by  
\begin{equation}
\epsilon_C=R, \;\;\;\;\;\; \epsilon_D=P.
\label{eq:epsilonCD}
\end{equation}

The measure of success we consider here is slightly different from the one considered in ref. \cite{PRE03}: 
To measure his success each player compares his profit $U_k(t)$ with the maximum between his {\em estimate} $\epsilon_k(t)$, given by (\ref{eq:epsilon}),
 and the arithmetic mean of the four payoffs given by
$\mu\equiv(R+S+T+P)/4$ \footnote{The reason to include the mean $\mu$ is to 
cover a wider range of situations than the ones permitted by the so-called {\it Pavlov's} rule. Pavlov strategy consists in to stick to the former move if it earned one of the two highest payoff but to switch in the contrary case. The measure considered here reduces to it when $R>\mu>P$.}. 
If $U^{R S T P}_k(t) \geq$ ($<$) 
$\max \{ \epsilon^{R S T P}_k, \mu \}$ the player assumes 
he is doing well (badly) and he keeps (changes) his $c_k(t)$
as follows: if player $k$ did well he assumes his $c_k(t)$ 
is adequate and he keeps it. On the other hand, if he did badly he 
assumes his $c_k$ is inadequate and he changes it (from C to D or from D 
to C in the deterministic version).

We are interested in measuring the average probability of 
cooperation $c$ vs. time, and in particular in its value of equilibrium $c_{eq}$, after a transient which is equivalent to the final 
fraction of C-agents $f_C$.

\section{Computation of the Equilibrium States}

\subsection{Deterministic version}

For the deterministic case the values of $c_{eq}$ are obtained by elementary calculus as follows.
Once equilibrium has been reached, the transitions from D to C, on average, must equal those from C to D.  
Thus, the average probability of cooperation $c_{eq}$ is obtained by equalizing the flux from C to D, $J_{CD}$, to the flux from D to C, $J_{DC}$. 
The players who play C either they get $R$ (in [C,C] encounters) or $S$ (in [C,D] encounters), and their 
estimate is $\epsilon_C = R$; thus, according to the update rule, they change to D if $R<\mu$ or $S<\max\{R,\mu\}$ respectively. 
For a given average probability of cooperation $c$, [C,C] encounters 
occur with probability $c^2$ and [C,D] encounters with probability $c(1-c)$. 
Consequently, $J_{CD}$ can be written as:
\begin{equation}
J_{CD} \propto a_{CC}c^2+a_{CD}c(1-c),
\label{eq:JCD}
\end{equation}
with 
\begin{equation}\label{eq:accacd}
\begin{array}{ccc}
 a_{CC}=\theta(\mu-R)\quad& \mbox{and} & \quad a_{CD}=\theta(\max\{R,\mu \}-S),
\end{array} 
\end{equation} 
where $\theta(x)$ is the step function given by:

\begin{equation}
\theta(x)\,=\,
\hspace{2mm}
\begin{array}{l}
 \,1 \quad if \quad x \,\geq0\quad\\ 
 \,0 \quad if \quad x \,<0\quad
\end{array} 
\end{equation}

On the other hand, the players who play D either they get $T$ (in [D,C] 
encounters) or $P$ (in [D,D] encounters) and their 
estimate is $\epsilon_D = P$; thus, according to the update rule, they change 
to C if $T<\max \{ \mu,P \}$ or $P<\mu$ respectively. As [D,C] encounters 
occur with probability $(1-c)c$ and [D,D] encounters with probability $(1-c)^2$, 
$J_{CD}$ can be written as:
\begin{equation}
J_{DC} \propto a_{DC}(1-c)c+a_{DD}(1-c)^2,
\label{eq:JDC}
\end{equation}
with 
\begin{equation}\label{eq:addadc}
\begin{array}{ccc}
 a_{DD}=\theta(\mu-P)\quad& \mbox{and} & \quad a_{DC}=\theta(\max\{P,\mu \}-T).
\end{array} 
\end{equation} 
In equilibrium 
\begin{equation}\label{eq:eqfluxes}
J_{CD}(c_{eq})=J_{DC}(c_{eq}),
\end{equation} 
and thus we get a set of second order algebraic equations for $c_{eq}$:
\vspace{1mm}
\begin{equation}
(a_{CC}-a_{CD} + a_{DC} - a_{DD})c_{eq}^2+ (a_{CD}-a_{DC} + 2a_{DD}) c_{eq}-a_{DD}=0.
\label{eq:eq-for-p}
\end{equation}
As there are 2 possibilities for each coefficient $a_{XY}$, we have a total 
of $2^4=16$ different equations governing all the possible equilibrium states
(actually there are 15 since this includes the trivial equation $0\equiv0$). 
The roots\footnote{The real roots $\in [0,1]$} of these equations are:\\

\begin{equation}\label{eq:roots1}
\begin{array}{l}
\vspace{1mm}
0\\\vspace{1mm}
\frac{3-\sqrt{5}}{2}\\\vspace{1mm}
1/2 \\\vspace{1mm}
\frac{\sqrt{5}-1}{2}\\
1
\end{array}
\end{equation}

In addition, we have to take into account the case when:
\begin{equation}
\begin{array}{l}
a_{CC}=a_{DD}=0 \\ 
a_{CD}=a_{DC}=1.
\end{array} 
\end{equation} 
In this case we can see from (\ref{eq:JCD}) and (\ref{eq:JDC}) that 
$J_{CD}\equiv J_{DC}$ \textit{identically}, so we have that 
$p_{eq}\equiv c_{o}$, (being $c_{o}$ the initial mean probability), 
whatever the initial conditions are.\\

For instance, for the canonical payoff matrix we have $a_{CC}=0=a_{DC}$
and $a_{CD}=1=a_{DD}$, therefore we get
\begin{equation}
c_{eq}(1-c_{eq})=(1-c_{eq})^2,
\label{eq:algebraic3051}
\end{equation}
with the root $c_{eq}=1/2$ corresponding to the stable dynamic equilibrium in 
which the agents change their state in such a way that, on average, half of the transitions are from C to D and the other half from D to C.

\subsection{ Stochastic version}

In the case of a continuous probability of cooperation $c_k$ 
, the calculation is a little bit more subtle: now the estimate $\epsilon_k$ for the agent $k$ is not only R or P, as it happened in the discrete case, but it can take a continuum of values as the probability $c_{k}$ varies in the interval [0,1]. From now on we will use the estimate as given in (\ref{eq:epsilon}), but instead of a $\epsilon_{k}$ as a function of time we will use a generic $\epsilon$  that is a function of the cooperation probability (and implicitly of time, off course), that is:
\begin{equation}
\epsilon^{R S T P}(c) =  (R-S-T+P) c(t)^2 +(S+T-2P) c(t) +P.
\label{eq:epsilon2}
\end{equation}
So we have:
\begin{equation}
\epsilon^{R S T P}_k(t) = \epsilon^{R S T P}(c_{k}(t)).
\end{equation}\\
To calculate $c_{eq}$ we begin by writing a balance equation for the probability $c_i(t)$. 
The agents will follow the same rule as before: they will keep their state if they are 
doing well (in the sense explained earlier) and otherwise they will change it. 
If two agents $i$ and $j$ play at time $t$, with probabilities $c_{i}(t)$ 
and $c_{j}(t)$ respectively, 
then the change in the probability $c_{i}$, provided he knows $c_{j}(t)$,
would be given by:
\begin{equation}\label{eq:updateci}
\begin{array}{ll}
\vspace{2mm}
c_{i}(t+1) - c_{i}(t) =  & - c_{i}(t)c_{j}(t)\,[1 - \theta(R - \epsilon^{R S T P}(c_{i}(t))\,\theta(R-\mu)]  \\ \vspace{2mm}
 & - c_{i}(t)[1-c_{j}(t)]\,[1 - \theta(S - \epsilon^{R S T P}(c_{i}(t))\,\theta(S-\mu)] \\ \vspace{2mm}
 & + [1-c_{i}(t)]c_{j}(t)\,[1 - \theta(T - \epsilon^{R S T P}(c_{i}(t))\,\theta(T-\mu)] \\ 
 & + [1-c_{i}(t)][1-c_{j}(t)]\,[1 - \theta(P - \epsilon^{R S T P}(c_{i}(t))\,
\theta(P-\mu)],
\end{array} 
\end{equation} 
being $\theta$ the step function.
The equation of evolution for $c_j(t)$
is obtained by simply exchanging $i\, \longleftrightarrow\,j$ in equation (\ref{eq:updateci}). 
Certainly, the assumption that each agent knows the probability of cooperation of
his opponent is not realistic. Later, when we perform the simulations, we will introduce a
procedure to estimate the opponent's probability  (more on this in Section {\bf
V.b})\\

In (\ref{eq:updateci}) if at time $t$ the payoff obtained by agent $i$, 
$X \, (\,= \,R,\, S, \,T\,\,$ or $\,P)$ is less than $\max \{ \epsilon^{R S T
P}(c_{i}(t)), \mu \}$, the first two terms in the RHS decrease the cooperation probability of agent $i$, while the two 
last terms increase it. The terms give no contribution if the payoff $X$ is greater or equal than $\max \{ \epsilon^{R S T P}(c_{i}(t)), \mu \}$.\\

We will use the canonical payoff matrix $M^{3051}$ to illustrate  how the above equation of evolution for $c_i(t)$ works. In this case, the estimate function is, by (\ref{eq:epsilon2}):
\begin{equation}
\epsilon^{3 0 5 1}(c)=-c^{2}+3c+1\;,
\label{eq:epsilon3051}
\end{equation}  
thus it is easy to see that:
\begin{equation}\label{eq:tita3051epsilon}
\begin{array}{ll}
\vspace{1mm}
\theta(3-\epsilon^{3 0 5 1}(c))\,=\,1 \quad & \forall \,c\, \in \,[0,1] \\ \vspace{1mm}
\theta(0-\epsilon^{3 0 5 1}(c))\,=\,0 \quad & \forall \,c\, \in \,[0,1] \\ \vspace{1mm}
\theta(5-\epsilon^{3 0 5 1}(c))\,=\,1 \quad & \forall \,c\, \in \,[0,1] \\ 
\theta(1-\epsilon^{3 0 5 1}(c))\,=\,0 \quad & \forall \,c\, \in \,(0,1]. 
\end{array}
\end{equation}\\
In addition we have for this case $\mu=2,25$, thus:
\begin{equation}\label{eq:tita3051mu}
\begin{array}{l}
\vspace{1mm}
\theta(3-\mu)\,=\,1\\\vspace{1mm}
\theta(0-\mu)\,=\,0\\\vspace{1mm}
\theta(5-\mu)\,=\,1\\ 
\theta(1-\mu)\,=\,0. 
\end{array}
\end{equation}\\
We can then write, to a very good approximation (we are assuming that the last 
line of (\ref{eq:tita3051epsilon}) is valid for $c=0$ also):
\begin{equation}\label{eq:updateci3051}
\begin{array}{ll}
\vspace{1mm}
c_{i}(t+1)-c_{i}(t) & = -c_{i}(t)[1-c_{j}(t)]+[1-c_{i}(t)][1-c_{j}(t)] \\ 
 & = [1-c_{j}(t)][1-2c_{i}(t)].
\end{array} 
\quad\forall i \neq j
\end{equation} 
Defining the mean probability of cooperation as
\begin{equation}
c=\frac{1}{N_{ag}}\sum_{i=1}^{N_{ag}}c_{i},
\end{equation}\\
summing eq. (\ref{eq:updateci3051}) over $i$ and $j$ leads to:
\begin{equation}\label{eq:meanci3051}
\begin{array}{ll}
\vspace{1mm}
c(t+1)-c(t) &  
= [1-c(t)][1-2 c(t)] \\ 
 &= 1 - 3 c(t) + c(t)^{2},
\end{array} 
\end{equation}\\
within an error of $O(1/N_{ag})$ since (\ref{eq:updateci3051}) is valid $\forall\, i \neq j$ but we are summing over all the $N_{ag}$ agents.\\
\\
Thereof we can calculate the equilibrium mean probability of cooperation $c_{eq}$:
\begin{equation}
0 = 1 - 3 c(t) + c(t)^{2},
\end{equation} 
obtaining the two roots:
\begin{equation}
c_{eq}\,=\,
\hspace{2mm}
\begin{array}{l}
\,1\quad\\ 
\,1/2\quad
\end{array} 
\end{equation}\\
being $c_{eq}=1/2$ the stable solution. Hence we obtain the same result that in the
deterministic case \\

Using analog reasoning for the general case, we can conclude that if 
\begin{equation}\label{eq:xnotin}
X\notin [\mu,\epsilon_{max}^{R S T P}]
\end{equation} 
or 
\begin{equation}\label{eq:mugteps}
\mu > \epsilon_{max}^{R S T P}
\end{equation} 
the results for the mean cooperation probability for the deterministic version and the
stochastic version, are the same.\\

There is an easy way to evaluate $\epsilon^{R S T P}_{max}$ in practice. 
It can be seen -see appendix- that \\
if
\begin{equation}\label{eq:epsmax1}
\begin{array}{ccc}
S+T>2 \max\{R,P\} \quad & \Rightarrow & \epsilon^{R S T P}_{max}=P- \frac{1}{4}\,\frac{(S+T-2P)^{2}}{(R-S-T+P)}
\end{array} 
\end{equation}\\
while, if
\begin{equation}\label{eq:epsmax2}
\begin{array}{ccc}
S+T\leq2 \max\{R,P\} \quad & \Rightarrow & \epsilon^{R S T P}_{max}=\max\{R, P\} \;.
\end{array} 
\end{equation} \\

When there is a payoff $X$ such that 
\begin{equation}\label{eq:xinto}
X\in [\mu,\epsilon_{max}^{R S T P}]
\end{equation} 
things can change because
agents who get $X$ update in general their probability of cooperation
$c_i(t)$ differently depending whether $X<\epsilon(c_i)$ or $X\geq\epsilon(c_i)$. So as the probability takes 
different values in the interval $[0,1]$, we have different equations of evolution, which somehow "compete" 
against each other in order to reach the equilibrium. The different equations that can appear are off course 
restricted to the ones generated by the coefficients $a_{XY}$ as they appear in (\ref{eq:eq-for-p}). It is 
reasonable to expect then that the final equilibrium value for the mean probability will be somewhere in between 
the original equilibrium values for the equations competing. We will analyze
some particular cases of this type in Section {\bf V.b} to illustrate this
point.\\

Although at first sight one may think that the universe of possibilities 
fulfilling condition (\ref{eq:xinto})
is very vast, it happens that no more than three different balance 
equations can coexist. This can be seen as follows: 
from eqs. (\ref{eq:epsmax1}) and (\ref{eq:epsmax2}), $\epsilon^{R S T P}_{max} \geq \max\{R, P\}$, and besides we 
know that the estimate never could be greater than all the payoffs, so there is at least one $X$ such that 
$\epsilon^{R S T P}_{max} < X$. So this leaves us with only two payoffs that effectively can be between $\mu$ 
and $\epsilon^{R S T P}_{max}$, and this results in at most three balance equations
playing in a given game.

\section{An example of coexistence of agents using different payoff  matrices: cooperation in presence of "always D" agents} 

Let us analyze now the situation where there are a mixing of agents using two 
different payoff matrices, each leading by separate to a different value 
of $c_{eq}$.  For simplicity we consider the deterministic version but the 
results for the stochastic version are similar.      
We call "antisocial" individuals those for whom cooperation 
never pays and thus, although they can initially be in the C state, after
playing they turn to state D and remain forever in this state. 
They can be represented by players using a payoff matrix that always
update $c_i$ to 0; for instance M$^{1053}$.
Notice that these individuals are basically different from those which use 
a payoff matrix fulfilling conditions (\ref{eq:inequal}) and 
(\ref{eq:dilemma2}) who, even though they realize the value of 
cooperation i.e. $R>P$ and $2R > T+S$, often may be tempted to "free ride"
in order to get a higher payoff. However, with the proposed mechanism -which implies
a sort of indirect reciprocity- when D grows above 50 \%  it punishes, 
on average, 
this behavior more than C favoring thus a net flux from D to C. Conversely, 
if C grows above 50 \% it punish, on average, this behavior more than D 
favoring thus the opposite flux from C to D. In other words,  
small oscillations around $f_C=0.5$ occur. On the other hand, agents using 
$M^{1053}$ are "immune" to the 
former regulating mechanism. Let us analyze the effect they have on 
cooperation when they "contaminate" a 
population of neutral agents (using the canonical payoff matrix).
In short, the two types of individuals play different games (specified by
different payoff matrices) without knowing this fact, a situation which does 
not seem too far from real life. 

The asymptotic average probabilities of cooperation  
can be obtained by simple algebra combining the update rules for 
M$^{3051}$ and M$^{1053}$.
The computation is completely analogous the one which leads to 
(\ref{eq:algebraic3051}). 
We have to calculate $J_{DC}$ and $J_{CD}$ as a function of the variable 
$c$ and the parameter $d$ and by equalizing them at equilibrium we 
get the equation for $c_{eq}$.
To $J_{DC}$ only contribute the fraction (1-$d$) of normal players using 
the canonical payoff matrix who play D against a player who also plays D 
(normal or antisocial). That is, $J_{DC}$ is given by
\begin{equation}
J_{DC}\propto (1-d) (1-c)^2.
\label{eq:JDC2}
\end{equation}
On the other hand, contributions to $J_{CD}$ come from one of these 3 types of encounters: 
i) [C,D] no matter if agents are neutral or antisocial, ii) [C,C] of two 
antisocial agents and iii) [C,C] of a neutral and antisocial  
agent (the neutral agent remains C and the antisocial, who started at $t=0$ 
playing C and has not played yet, changes from C to D). 
The respective weights of these 3 contributions are: $c(1-c)$, 
$d^2c^2$ and $\frac{1}{2}2d(1-d)c^2$. Therefore, $J_{CD}$ is given by  
\begin{equation}
J_{CD}\propto c(1-c)+ d^2c^2+d(1-d)c^2=c(1-c)+dc^2.
\label{eq:JCD2}
\end{equation}
In equilibrium $J_{DC}=J_{CD}$ and the following equation 
for $c_{eq}$ arises:
\begin{equation}
(1-d)( 2c^2_{eq}-2c_{eq}+1)+c_{eq}=0,
\label{eq:algebraic3051-1053}
\end{equation}
and solving it: 
\begin{equation}
c_{eq}=\frac{3-2d \pm \sqrt{-4d^2+4d+1}}{4(1-d)}.
\label{eq:roots}
\end{equation}
We must take the roots with the "-" sign because those with "+" are 
greater than 1 for non null values of d. We thus get the following 
table for $c_{eq}$ for different values of the parameter $d$: 

\begin{center}

\begin{tabular}{|p{3.5cm}|}
\hline
$c_{eq}$ (d=0.0) = 0.5000\\
\hline
$c_{eq}$ (d=0.1) = 0.4538\\
\hline
$c_{eq}$ (d=0.2) = 0.4123\\
\hline
$c_{eq}$ (d=0.3) = 0.3727\\
\hline
$c_{eq}$ (d=0.4) = 0.3333\\
\hline
$c_{eq}$ (d=0.5) = 0.2929\\
\hline
$c_{eq}$ (d=0.6) = 0.2500\\
\hline
$c_{eq}$ (d=0.7) = 0.2029\\
\hline
$c_{eq}$ (d=0.8) = 0.1492\\
\hline
$c_{eq}$ (d=0.9) = 0.0845\\
\hline
$c_{eq}$ (d=1.0) = 0.0\\
\hline
\end{tabular}

\end{center}

\vspace{1mm}

Table 1. $c_{eq}$ for agents using M$^{3051}$ contaminated
by a fraction $d$ of antisocial agents using  M$^{1053}$.\\

\section{Simulations}

\subsection{Deterministic version}

In this subsection we present some results produced by simulation for the
deterministic version. Different payoff matrices were simulated and it was found that the 
system self-organizes, after a transient,
in equilibrium states in total agreement with those calculated in 
(\ref{eq:roots1}). 

The update from $c_i(t)$ to $c_i(t+1)$ was dictated by a balance 
equation of the kind of (\ref{eq:updateci}).
The measures are performed over 1000 simulations each
and $\bar{c}_{eq}$ denotes the average of $c_{eq}$ over these milliard of
experiments.
In order to show the independence from the initial distribution of 
probabilities of cooperation, Fig. 1 shows the evolution with time of the 
average probability of cooperation for different initial proportions of 
C-agents $f_{C0}$ for the case of the canonical payoff matrix M$^{3051}$ 
(i.e. $R=3, S=0, T=5$ and $P=1$).
\begin{center}
\begin{figure}[h]
\centering
\psfig{figure=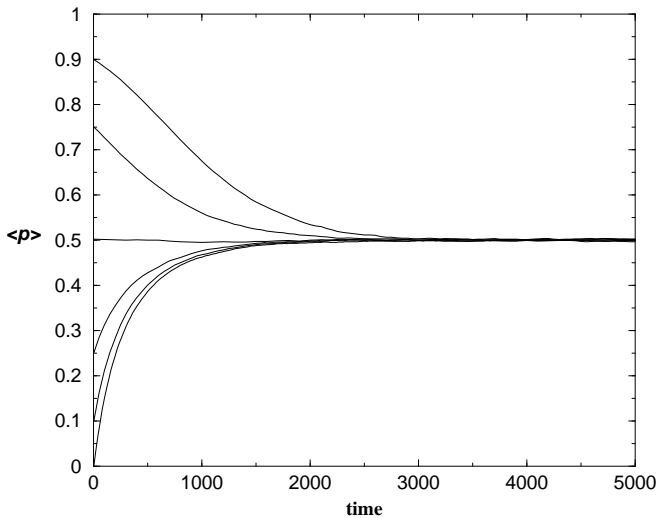,height=7cm}
\caption{ $\bar{c}$ vs. time, for different initial values of 
$f_{C0}$, for the canonical payoff matrix.} 
\end{figure}
\end{center}

Depending on the payoff matrix the equilibrium asymptotic states can be of 
three types: of "all C" ($\bar{c}_{eq}$ = 1.0), "all D" ($\bar{c}_{eq}$ = 0.0)
or something in between ($0<\bar{c}_{eq}<1$).

We have seen that the canonical payoff matrix M$^{3051}$ provides an example of matrix which gives $\bar{c}_{eq} =0.5$. 

Let us see examples of payoff matrices which produce other values of $\bar{c}_{eq}$.
A payoff matrix which produces $\bar{c}_{eq} =1.0$ is obtained simply by permuting the canonical values of $S$ (0) and $T$ (5), i.e. $M^{3501}$.
For this matrix we have, by inspection of (\ref{eq:accacd}) and (\ref{eq:addadc}): 
\begin{equation}
\begin{array}{cc}
a_{CC}=a_{CD}=0 \quad & \quad a_{DC}=a_{DD}=1.
\end{array} 
\end{equation} 
Hence, after playing the PD game the pair of agents always ends [C,C] since $J_{CD}\equiv 0$ by (\ref{eq:JCD}).

On the other hand, a payoff matrix which leads $\bar{c}_{eq} =0.0$ is 
obtained simply by permuting the canonical values of $R$ (3) and $P$ (1), i.e.
$M^{1053}$, for which:
\begin{equation}
\begin{array}{cc}
a_{CC}=a_{CD}=1 \quad & \quad a_{DC}=a_{DD}=0.
\end{array} 
\end{equation} 
That is, all the changes are from C to D since in this case $J_{DC}\equiv 0$

The rate of convergence to the possible values of $\bar{c}_{eq}$ depends on the 
values of $J_{CD}$ and $J_{DC}$. 

Fig. 2 shows the approach of the average probability of cooperation for different 
payoff matrices to their final 5 equilibrium values. 
\begin{center}
\begin{figure}[h]
\centering
\psfig{figure=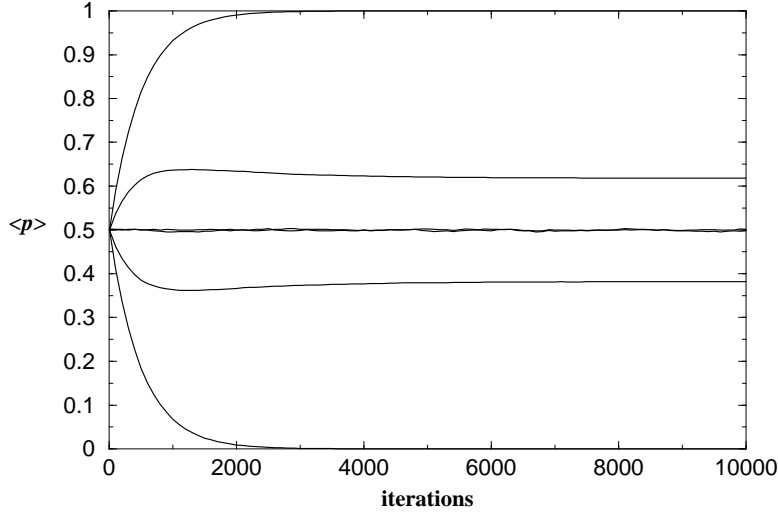,height=7cm}
\caption{Curves of $\bar{c}$ vs. time for different payoff matrices 
producing the 5 possible values of $c_{eq}$ 
(from below to above): payoff matrices M$^{3501}$
with $c_{eq}=1$, M$^{2091}$ with $c_{eq}\simeq 0.62$,
 M$^{3051}$ with $c_{eq} = 0.5$, M$^{2901}$ with $c_{eq}
\simeq 0.38$ and M$^{1035}$ with $c_{eq}= 0$.} 
\end{figure}
\end{center}

Finally, we simulated the mixing of agents using payoff matrices      
M$^{3051}$ and M$^{1053}$.
The evolution to equilibrium states for different fixed fractions $d$ 
of agents using M$^{1053}$ is presented in Fig. 3. The results are in complete agreement with the asymptotic probabilities of cooperation which appear in Table 1. 
\begin{center}
\begin{figure}[h]
\centering
\psfig{figure=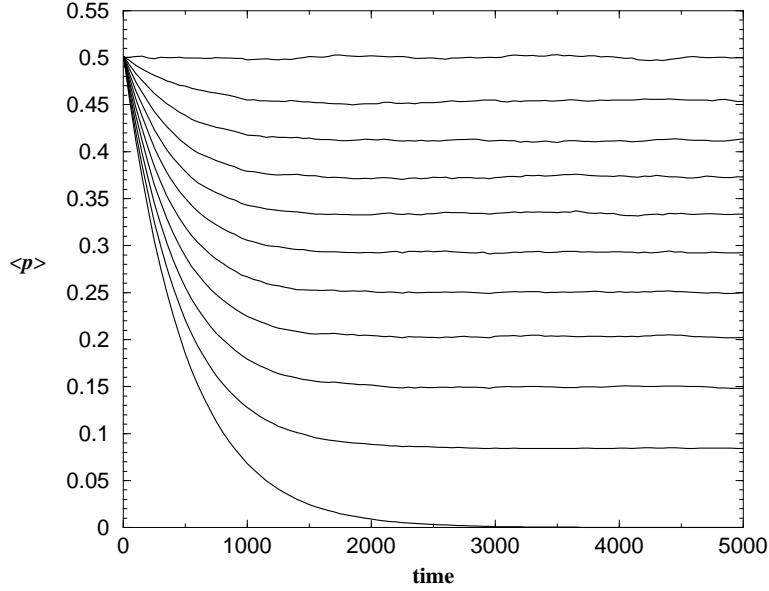,height=8cm}
\caption{ The evolution of $\bar{c}$ with time, for different 
values of the fraction $d$ "antisocial" agents (using M$^{1053}$) embedded in 
a population of neutral agents (using the canonical payoff matrix).} 
\end{figure}
\end{center}

\subsection{ Stochastic version}

In this case simulations were made updating the probability of cooperation 
according to eq. (\ref{eq:updateci}). However, as we 
anticipated, we have to change slightly this eq. to reflect reality: 
two agents $i$ and $j$ interact and they obtain the payoffs $X_{i}$ 
and $X_{j}$, respectively. 
For each of them there is no way, from this only event, to know the
probability of cooperation $c_k$ 
of his opponent. What they can do then is to (roughly) estimate 
this $c_k$ as follows. The player $i$ average utility in an encounter at 
time $t$ with agent $j$ is given by:
\begin{equation}\label{eq:uti}
U_{ij}(t) = R\,c_{i}(t)c_{j}(t) + S\,c_{i}(t)[1 - c_{j}] + T\,[1 - c_{i}(t)]c_{j}(t) + P\,[1 - c_{i}(t)][1 -
c_{j}(t)] \quad .
\end{equation} 
When he plays he gets the payoff $X_{i}$, so his best estimate
$\tilde{c}_{j}^{i}$ for the probability of agent $j$ is obtained by
replacing $U_{ij}(t)$ for $X_{i}$ in eq. (\ref{eq:uti}). Then he will have:
\begin{equation}\label{eq:estcj}
\tilde{c}_{j}^{i}(t) = \frac{X_{i} - P + c{i}(t)(P - S)}{c_{i}(t)(R - S -T -P) +
T - P}
\end{equation}
Exchanging $i$ for $j$ in this eq. gives the estimate of the probability
$c_{i}(t)$ that makes agent $j$. Equation (\ref{eq:estcj}) can retrieve any
value of $\tilde{c}_{j}^{i}(t)$ and not just in the interval $[0, 1]$, so it is
necessary to make the following replacements:
\begin{equation}\label{eq:truncation}
\begin{array}{cccc}
\mbox{if} \quad & \tilde{c}_{j}^{i}(t)>1 \quad & \Longrightarrow & \quad \tilde{c}_{j}^{i}(t)=1 \\
\mbox{and if} \quad & \tilde{c}_{j}^{i}(t)<0 \quad & \Longrightarrow & \quad \tilde{c}_{j}^{i}(t)=0
\end{array}
\end{equation}
When this happens, the agent is making the roughest approximation, which is
to assume that the other player acts like in the deterministic case.\\

For the canonical payoff matrix, the result was the expected one as this is a 
matrix obeying condition (\ref{eq:xnotin}): as predicted by the analytical 
calculation of Section {\bf III.b}, the value for the equilibrium mean 
probability is $c_{eq}=1/2$ as in the deterministic case, despite the change introduced in (\ref{eq:estcj}). 
Simulations for other payoff matrices satisfying
conditions (\ref{eq:xnotin}) or (\ref{eq:mugteps}) were also made and in all
the cases the deterministic results were recovered.\\

We will illustrate the case in which some
\begin{equation}\label{eq:xin}
X\in [\mu,\epsilon_{max}^{R S T P}]
\end{equation} 
with two particular examples. One of them is the case of the 
normalized matrix $M^{1 S 1 0}$, with $S$ varying from $1$ to $2$, both 
limiting cases in which condition (\ref{eq:xin}) ceases to be valid. 
So for $S\leq1$ the update equation is given simply by:
\begin{equation}\label{eq:sle1}
c_{i}(t+1) - c_{i}(t) = [1 - c_{i}(t)] [1 - c_{j}(t)]
\end{equation} 
being $c_{eq}=1$ in this case, while for $S>2$:
\begin{equation}\label{eq:sgt2}
c_{i}(t+1) - c_{i}(t) = 1 - c_{i}(t) - c_{i}(t)c_{j}(t)
\end{equation}
for which $c_{eq}= \frac{\sqrt{5}-1}{2}$ is the corresponding equilibrium value.
When $S \in (1,2]$, both balance equations play a role, the 
general equation for the update follows from eq. (\ref{eq:updateci})
applied to this particular case:
\begin{equation}\label{eq:1s2}
c_{i}(t+1) - c_{i}(t) = [1 - c_{i}(t)] [1 - c_{j}(t)] + [c_{j}(t) - 
2 c_{i}(t)c_{j}(t)][1 - \theta(1 -\epsilon)].
\end{equation} 
So we can see that for $R=T=1 \geq \epsilon$, eq. (\ref{eq:1s2}) reduces to 
(\ref{eq:sle1}) while if $R=T=1 < \epsilon$ we obtain (\ref{eq:sgt2}). When
the simulation takes place, $c_{j}$ has to be replaced by
$\tilde{c}_{j}^{i}$. \\

The same analysis can be done for the matrices $M^{1 1 T 0}$, with $T$
varying from $1$ to $2$ also. In this case the other root competing with
$c_{eq}=1$ is $c_{eq}= \frac{3 - \sqrt{5}}{2}$.\\

The results of the simulations for both cases are presented in the next
table, data for $S>2$ and $T>2$ -for which (\ref{eq:xnotin}) is valid- is
also included \footnote{$\bar{c}_{eq}$ corresponds to the average of $c_{eq}$ over 100 experiments.}:

$$\begin{tabular}{|p{1.5cm}|p{1.5cm}|p{1.5cm}|}
\hline $X$ & $\bar{c}_{eq}$ for $X=S$  & $\bar{c}_{eq}$ for $X=T$ \\ 
\hline 1 & 1 & 1\\    
\hline 1.5 & 1 & 1\\    
\hline 1.9 & 1 & 1\\ 
\hline 2 & 1 & 1\\
\hline 2.1 & 0.617 & 0.383\\ 
\hline 4 & 0.581 & 0.370\\
\hline 8 & 0.556 & 0.403\\
\hline 16 & 0.530 & 0.467\\
\hline 1000 & 0.548 & 0.455\\
\hline 
\end{tabular} $$

As it can be seen from the data, for $1 <  X \leq 2$, that is,
when condition (\ref{eq:xin}) is valid, the results for the stochastic case
are the same that they would be if we were working with the deterministic
model. This is a consequence of the estimate (\ref{eq:estcj}) together with
conditions (\ref{eq:truncation}). \\

For values of $T$ and $S$ greater than $2$, for which condition
(\ref{eq:xin}) does not hold any more, we can observe what at first may seem
a curiosity: for $T$ or $S$ near $2$, the equilibrium values for the
deterministic case are recovered as expected, but as we increase the values
of $T$ or $S$, the value of $c_{eq}=1/2$ is approached. After a little 
thought, it is clear that this is also a consecuence of the estimation 
of (\ref{eq:estcj}),
since it depends on the payoffs. It can be easily seen that in the case of
$M^{1 1 T 0}$:  
\begin{equation}\label{eq:tgg1}
\mbox{if} \quad T \gg 1  \quad \mbox{then} \quad \tilde{c}_j^{i} \simeq
0 \quad \forall \, i \, , \, j \quad \mbox{(for $X_{i} \neq T$)} .  
\end{equation}
If we take then $c_{j}=0$ in eq. (\ref{eq:updateci}), and remembering that $T \rightarrow \infty$ implies that $\mu \rightarrow \infty$, we will obtain that
$c_{eq}=1/2$. In an analogous way for $M^{1 S 1 0}$:
\begin{equation}\label{eq:sgg1}
\mbox{if} \quad S \gg 1 \quad \mbox{then} \quad \tilde{c}_j^{i} \simeq 1 
\quad \forall  \, i \, , \, j \quad \mbox{(for $X_{i} \neq S$)}  
\end{equation}
which toghether with eq. (\ref{eq:updateci}) again leads to $c_{eq}=1/2$. The encounters for which $X_{i} = S$ or $T$ are responsible for that the exact value $c_{eq}=1/2$ is not attained. A similar analysis can be done when $R$ or $P \rightarrow \infty$.

\section{SUMMARY AND OUTLOOK}

The proposed strategy, the combination of measure of success and 
update rule, produces cooperation for a wide
variety of payoff matrices. 

In particular, notice that: 

\begin{itemize}

\item A cooperative regime arises for payoff matrices representing \textit{"Social Dilemmas"} like the canonical one. 
On the other hand spatial game algorithms like the one of ref. \cite{nm93} produce cooperative states ($c_{eq}> 0$) in general for the case of a "weak dilemma" in which $P$ = $S$ = 0 or at most when $P$ is significantly below $R$ \footnote{In particular, in a spatial game in which each player interacts with his four nearest neighbors, we have checked that the canonical payoff matrix
lead to the an "all D" state with $c_{eq}=0$.}.

\item Payoff matrices with $R=S=0$ which, at least in
principle, one would bet that favor D, actually generate equilibrium
states with $c_{eq}\neq 0$, provided that $P<\mu$ -see eqs. (\ref{eq:JCD})-(\ref{eq:eqfluxes}).

\item Any value of equilibrium average cooperation can be reached 
in principle, even in the case of the deterministic model, by the appropriate 
mixing of agents using 2 different payoff matrices. 
This is an interesting result that goes beyond the different existent social 
settings. For instance we have in mind situations in which one wants to 
design a device or mechanism with a given value of $c_{eq}$ that optimizes 
its performance. 

\item In this work we adopted a {\it Mean Field} approach in which all the 
spatial correlations between agents were neglected. One 
virtue of this simplification is that it shows the model does not require
that agents interact only with those within some geographical proximity 
in order to sustain cooperation. Playing with fixed neighbors
is sometimes considered as an important ingredient to
successfully maintain the cooperative regime \cite{nm93},\cite{cra2001}.
(Additionally, the equilibrium points can be obtained by simple algebra.)

\end{itemize}

To conclude we mention different extensions and applications of this model as possible future work.
We mentioned, at the beginning, "statistical mechanic" studies.
For instance, by dividing the four payoffs between say the reward $R$ 
reduces the parameters to three: $a=S/R$, $b=T/R$ and $d=P/R$,
and we are interested to analyze the dependence of $c_{eq}$ on each one 
of these 3 parameters 
in the vicinity of a transition between two different values.
It is also interesting to introduce noise in the system, by means 
of an inverse temperature parameter $\beta$, in order to allow irrational choices.
The player $i$ changes his strategy with a probability $W_i$ given by  

$$W_i=\frac{1}{1+\exp[\beta(U_i-\tilde{\epsilon_i})]},$$

where $\tilde{\epsilon}_i\equiv \max\{\epsilon_i,\mu\}$.

We are planning also a quantum extension of the model in order to 
deal with players which use superposition of strategies 
$\alpha_C |C > + \alpha_D |D>$ instead of 
definite strategies.

The study of the spatial structured version and how the different agents lump together is also an interesting problem to consider. Results on that topic will be presented elsewhere.

Finally, a test for the model against experimental data seems interesting.
In the case of humans the experiments suggest, for a given class of games 
({\it i.e.} a definite rank in the order of the payoffs),  
a dependency of $f_c$ with the relative weights of $R,S,T$ and $P$, which is 
not observed in the present model.
Therefore, we should change the update rule in such a way to capture this 
Feature. Work is also in progress in that direction. \\

\vspace{3mm}

{\bf APPENDIX: Calculus for the maximum of the Gain Estimate function in the stochastic case.} 
\vspace{2mm}

We will now show in detail the calculus for the maximum of the gain estimate function $\epsilon^{R S T P}(c)$, restricted to the interval $[0,1]$.
First we have to know if the function has a maximum in the open interval $(0,1)$. This can be done by noticing that, by (\ref{eq:epsilon2}), for having negative concavity, we have the condition:
\begin{equation}\label{eq:epsmax}
R-S-T+P<0
\end{equation}
By doing 
\begin{equation}
\frac{d}{dc}\epsilon^{R S T P}=0
\end{equation}
we find that the extremum of $\epsilon^{R S T P}(c)$ is attained at
\begin{equation}
c_{o}=-\frac{1}{2}\,\frac{(S+T-2P)}{(R-S-T+P)}
\end{equation}  
Imposing $c_{o}>0$, $c_{o}<1$ and using (\ref{eq:epsmax}) for consistency, we obtain:
\begin{equation}\label{eq:cond}
\begin{array}{ll}
S+T>2P\quad &,\quad S+T>2R 
\end{array} 
\end{equation} 
Notice that the sum of this two conditions is equivalent to condition (\ref{eq:epsmax}). In turn, (\ref{eq:cond}) can be expressed as
\begin{equation}\label{eq:ineqrstp}
S+T>2 \max\{R,P\}
\end{equation}
so this inequality resumes (\ref{eq:epsmax}) and (\ref{eq:cond}). It can be seen that if (\ref{eq:ineqrstp}) is fulfilled,  $\epsilon^{R S T P}_{max}\geq\mu$ always.\\

So if condition (\ref{eq:ineqrstp}) holds, the maximum of the function $\epsilon^{R S T P}(c)$ takes place in the interval $(0,1)$ and its value as a function of the parameters $R, S, T$ and $P$ is:
\begin{equation}
\epsilon^{R S T P}_{max}=P- \frac{1}{4}\,\frac{(S+T-2P)^{2}}{(R-S-T+P)}
\end{equation}
On the other hand, if 
\begin{equation}\label{eq:ine2qrstp}
S+T\leq2\max\{R,P\}
\end{equation}
then
\begin{equation}
\epsilon^{R S T P}_{max}=\max\{R, P\}
\end{equation} 
since $\quad \epsilon^{R S T P}(0)=P\quad$, $\quad \epsilon^{R S T P}(1)=R$.


\vspace{2mm}


\begin{thebibliography}{99}

\bibitem{hs88} J. Hofbauer and K. Sigmund, {\it The Theory of Evolution 
and Dynamical Systems}, Cambridge University Press 1988.

\bibitem{vNM44} J. von Neumann and O. Morgenstern, {\it Theory of Games 
and Economic Behavior}, Princeton University Press, Princeton, 1944. 

\bibitem{ms} J. Maynard-Smith, {\it Evolution and the Theory of Games}, 
Cambridge Univ. Press 1982. 

\bibitem{tch99} P.E. Turner and L. Chao, Nature (London) {\bf 398},
441 (1999).

\bibitem{meyer99} D.A. Meyer, Phys. Rev Lett. {\bf 82}, 1052-1055 (1999).

\bibitem{ewl99}  J. Eisert, M. Wilkens and M. Lewenstein, Phys. Rev. Lett. {\bf 83} 3077-3080 (1999).

\bibitem{lj02} C.F. Lee and N. Johnson, arXiv.org/quant-ph/abs/0207012 (2002). 

\bibitem{lj02b} C.F. Lee and N. Johnson, Phys. Lett. {\bf A 301} 343-349; quant-ph/0207080 (2002). 

\bibitem{st99} G. Szab\'o and C. T\"oke, Phys. Rev. {\bf E 58}, 69 (1998).

\bibitem{sasd00} G. Szab\'o, T. Antal, P. Szab\'o and M. Droz, Phys. Rev. {\bf E 62}, 1095 (2000).

\bibitem{f52} M. Flood, "Some Experimental Games", Research Memorandum, RM-789-1, 20 June 1952, The RAND Corporation, 1700 Main St., Santa Monica, CA
 (1952).

\bibitem{n51} J. Nash, Annals of Mathematics {\bf 54}, 286 (1951).

\bibitem{rg66} A. Rapoport and M. Guyer, {\it General Systems} {\bf 11},
205 (1966).

\bibitem{axel84} R. Axelrod, {\it The Evolution of Cooperation}, 
Basic Books, New York, 1984.

\bibitem{ep98} J. Epstein, 
{\it Zones of Cooperation in Demographic Prisoner's Dilemma} 
Complexity, Vol. 4, Number 2, November-December 1998.  

\bibitem{nm93} M.A. Nowak and R. May, Int. J. Bifurcation and Chaos {\bf 3}, 35 (1993); M.A. Nowak and R. May, Nature {\bf 359}, 826 (1992).

\bibitem{PRE03} H. Fort, Phys. Rev. {\bf E 68 }, 026118 (2003).

\bibitem{W95} J. W. Weibull , {\it Evolutionary Game Theory},
MIT Press 1995. 

\bibitem{cra2001} M.D. Cohen, R.L. Riolo and R. Axelrod, 
Rationality and Society {\bf 13}, 5 (2001).





\end{thebibliography}
\end{document}